\documentclass[5p, authoryear]{elsarticle}

\usepackage{amsmath,amssymb,amsfonts}
\usepackage{algorithmic}
\usepackage{graphicx}
\usepackage{float}
\usepackage{algorithm}
\usepackage{eurosym}
\usepackage{booktabs}
\usepackage{multirow}
\usepackage{bm}
\usepackage[normalem]{ulem}

\PassOptionsToPackage{hyphens}{url}\usepackage{hyperref}

\makeatletter
\def\ps@pprintTitle{%
 \let\@oddhead\@empty
 \let\@evenhead\@empty
 \def\@oddfoot{}%
 \let\@evenfoot\@oddfoot}
\makeatother

\begin{document}

\begin{frontmatter}

\title{An Artificial Intelligence Solution for Electricity Procurement in Forward Markets}
\author[1]{Thibaut Th\'{e}ate}\corref{cor1}
\ead{thibaut.theate@uliege.be}
\author[1]{S\'{e}bastien Mathieu}
\ead{smathieu@uliege.be}
\author[1]{Damien Ernst}
\ead{dernst@uliege.be}
\address[1]{Montefiore Institute, University of Li\`{e}ge (All\'{e}e de la d\'{e}couverte 10, 4000 Li\`{e}ge, Belgium)}
\cortext[cor1]{Corresponding author}

\begin{abstract}

    Retailers and major consumers of electricity generally purchase an important percentage of their estimated electricity needs years ahead in the forward market. This long-term electricity procurement task consists of determining when to buy electricity so that the resulting energy cost is minimised, and the forecast consumption is covered. In this scientific article, the focus is set on a yearly base load product from the Belgian forward market, named calendar (CAL), which is tradable up to three years ahead of the delivery period. This research paper introduces a novel algorithm providing recommendations to either buy electricity now or wait for a future opportunity based on the history of CAL prices. This algorithm relies on deep learning forecasting techniques and on an indicator quantifying the deviation from a perfectly uniform reference procurement policy. On average, the proposed approach surpasses the benchmark procurement policies considered and achieves a reduction in costs of 1.65\% with respect to the perfectly uniform reference procurement policy achieving the mean electricity price. Moreover, in addition to automating the complex electricity procurement task, this algorithm demonstrates more consistent results throughout the years. Eventually, the generality of the solution presented makes it well suited for solving other commodity procurement problems.
    
\end{abstract}


\end{frontmatter}

\section{Introduction}
\label{SectionIntroduction}

Electricity retailers generally buy an important share of their consumption years ahead in the forward market. They have to accurately estimate their clients' consumption and purchase the appropriate quantity of electricity. This challenging task also applies to major electricity consumers which sign flexible bilateral contracts with their energy retailer. Typically, they have to decide when to purchase blocks of energy at a price generally indexed on the forward prices. Each block corresponds to a certain percentage of their total electricity consumption, this quantity being formerly predicted by the retailer. Eventually, the potential discrepancy between electricity purchased and forecast consumption is covered by the retailer at the end of the procurement horizon. The long-term electricity procurement problem consists of determining when to purchase electricity in the forward market, so that the predicted consumption is secured and the energy cost is minimised. This decision-making problem is particularly challenging because of its sequential and highly stochastic nature, coupled with a poorly observable environment.\\

Nowadays, the long-term electricity procurement task is generally performed by experienced consultants, based on customised rules and their expectations regarding the future energy market direction. This research paper proposes an alternative approach: an algorithm providing recommendations to either buy electricity now or to wait for a future opportunity, based on the history of forward prices. This solution may interest these consultants, but also retailers who are willing to deploy more advanced procurement techniques and major consumers choosing not to rely on consultants for buying their electricity.\\

The objective of this research is to develop a new simple yet efficient solution to be deployed in the industry to solve the complex sequential decision-making problem behind the long-term electricity procurement task. Although forecasting techniques are considered in the proposed algorithm, this work does not focus on improving the forecasts of forward prices. Instead, the main contribution of this research paper is related to the entire decision-making process: how to make relevant trading decisions based on imperfect information, including inaccurate forecasts.\\

The algorithm presented in this scientific article is based on the idea that the purchase decisions should be split over the procurement horizon to spread the trading risk, with a nominal anticipation or delay depending on the market direction. This algorithm relies on a forecasting mechanism to predict the dominant market trend, and on an indicator quantifying the deviation from a perfectly uniform reference procurement policy to trigger purchase decisions. In addition to classical approaches, deep learning (DL) techniques are considered for the forecasting task because deep neural networks (DNNs) are able to efficiently handle temporal dependence and structures like trends.\\

The present scientific research paper is structured as follows. To begin with, Section \ref{SectionMaterialsAndMethods} thoroughly presents the novel approach proposed to solve the long-term electricity procurement task. Firstly, a concise review of the scientific literature referred to in this problem is presented in Section \ref{SectionLiteratureReview}. Secondly, Section \ref{SectionProblemStatement} introduces a formalisation of this complex sequential decision-making problem. Thirdly, an algorithm is proposed in Section \ref{SectionAlgorithm} to solve the long-term electricity procurement task. Following on, Section \ref{SectionResults} describes the performance assessment methodology and presents the results achieved using this algorithm. Then, Section \ref{SectionDiscussion} provides an additional discussion about the proposed algorithmic solution and some suggestions for several avenues for future work. Eventually, Section \ref{SectionConclusions} concludes.

\section{Materials and Methods}
\label{SectionMaterialsAndMethods}

In this section, the novel algorithmic solution proposed to solve the sequential decision-making problem behind the long-term electricity procurement task is presented in detail. Section \ref{SectionLiteratureReview} analyses the scientific literature on this particular topic. Section \ref{SectionProblemStatement} then comprehensively formalises the sequential decision-making problem at hand. Subsequently, Section \ref{SectionAlgorithm} presents the algorithm developed to solve this problem.

\subsection{Literature review}
\label{SectionLiteratureReview}

The scientific literature proposes multiple strategies for electricity producers willing to sell their energy in the forward market. On the other hand, the sides of the retailers and consumers lack proper scientific coverage, with only a few articles currently available. The solutions presented are typically based on stochastic programming and optimisation techniques. Article \citep{Carrin2007} proposes a solution to the electricity procurement problem faced by a major consumer whose supply sources include bilateral contracts, self-production and the day-ahead market. A stochastic programming approach is considered, with risk aversion being modelled using the conditional value at risk (CVaR) methodology. The proposed solution is assessed through a realistic case study which highlights the trade-off between cost minimisation and risk mitigation. One chapter of the book \citep{Conejo2010} is dedicated to the electricity procurement problem from a major consumer perspective, while another chapter discusses the case of a retailer in a medium-term horizon. In both cases, the electricity procurement problem is mathematically formulated as a multi-stage stochastic programming problem, where the evolution of the price is modelled as a stochastic process using a set of scenarios and the risk aversion is modelled through the CVaR. The work concludes that multi-stage stochastic programming appears to be an appropriate modelling framework to make electricity procurement decisions under uncertainty, with the complex multi-stage stochastic model being translated into a tractable mixed-integer linear programming problem. Article \citep{Zare2010} introduces a technique based on information gap decision theory to assess different procurement strategies for major consumers. The objective is not to minimise the procurement cost but rather to assess the risk aversion of some procurement strategies with respect to the minimum achievable cost. The results suggest that strategies related to a higher procurement cost are more robust and risk averse. Article \citep{Nojavan2015} proposes a robust optimisation approach to solve the electricity procurement problem from a retailer perspective. A collection of robust mixed-integer linear programming problems is formulated, with the electricity price uncertainty being modelled by considering upper and lower limits for the energy prices rather than the forecast prices. Articles \citep{Beraldi2017Bis} and \citep{Beraldi2017} present a stochastic optimisation approach relying on the integration of the paradigm of joint chance constraints and the CVaR risk measure to solve the electricity procurement problem from a consumer perspective. The results for a real case study highlight the trade-off between risk and reliability by considering different levels of risk aversion. Article \citep{Zhang2018} proposes another multi-stage stochastic programming model for the long-term electricity procurement problem faced by a major consumer, where the complexity of the task is reduced by dividing a one-year planning into seasons. In this model, a season is represented by characteristic weeks and the seasonal demand is revealed at the beginning of each season. Article \citep{Hu2018} presents a short-term decision-making model based on robust optimisation to help an electricity retailer in determining both the electricity procurement and its electricity retail price so that profit is maximised. Two possibilities are offered to the retailer for its electricity procurement task: directly purchasing energy from generation companies and buying electricity on the spot market. Article \citep{Konishi2018} tackles a slightly different aspect of the electricity procurement problem as it studies how to size and use energy storage systems to minimise the procurement costs of electricity. The study focuses on short-term energy procurement by considering both the day-ahead market and the real-time market. Article \citep{Wang2019} studies a multi-period electricity procurement problem in the specific context of smart-grid communities. The required energy can be obtained from both the day-ahead market, characterised by variable prices, and renewable energy sources which are free but with uncertain supplies. To determine the optimal procurement amount, the authors consider an approach based on dynamic programming which has been proven to provide significant cost-savings. Article \citep{Zhang2019} introduces an agent-based two-stage trading model for direct electricity procurement of major consumers, which considers both the fairness and efficiency of direct energy procurement. According to the authors, this novel mechanism could offer more choices for both major consumers and generation companies which could also benefit from the reduction of the average market price.\\

To summarise, the solutions to the electricity procurement problem presented in existing scientific literature are mainly based on stochastic programming, dynamic programming and optimisation techniques. Moreover, the sequential decision-making problem behind the electricity procurement task is formalised in many different ways (time horizons, electricity power sources and markets, electricity consumption), meaning that a fair comparison between these solutions is not really feasible. Nevertheless, despite being very sound and interesting works, these approaches are not well established on a large scale within the industry. One reason for this is certainly the fact that these solutions are generally black box models which can be quite difficult for inexperienced employees to understand, interpret and monitor on a daily basis. Another observation regarding the scientific literature is the surprising absence of approaches based on advanced artificial intelligence techniques. To fill this gap, the present research paper introduces a novel algorithmic solution taking advantage of the recent and promising results achieved by DL techniques in many fields. Therefore, the main contribution of this research work is as follows: the development of an algorithm building on the promise of DL techniques to provide accurate forecasts to make sound and explicable trading decisions for long-term electricity procurement.

\subsection{Problem formalisation}
\label{SectionProblemStatement}

\begin{figure}
    \centering
    \includegraphics[scale=0.52, trim={1.1cm 1.5cm 2cm 1.5cm}, clip]{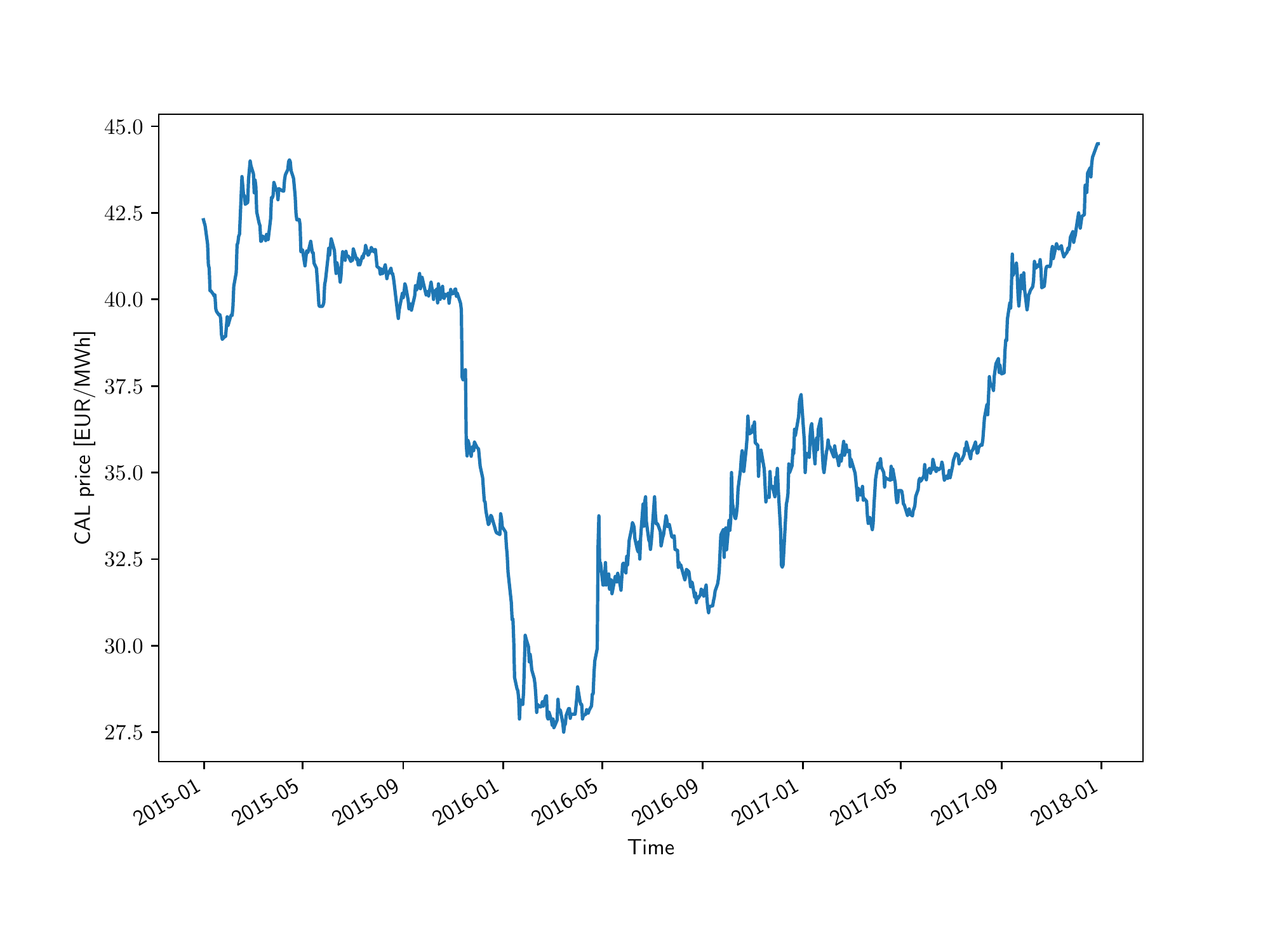}
    \caption{Illustration of the CAL 2018 product}
    \label{CALIllustration}
\end{figure}

In this section, the long-term electricity procurement problem considered is thoroughly presented and formalised. It is assumed that the only supply source at the disposal of the agent, whether a retailer or a consumer, is the calendar product (CAL) from the Belgian forward market operated by Ice Endex (Belgian Power Base Futures). This yearly base load product is tradable up to three years ahead of the delivery period. For instance, the CAL 2018 product corresponds to the delivery of electricity for the entire year 2018, this energy being tradable between 2015 and 2017 included, as depicted in Figure \ref{CALIllustration}. This is a slight simplification of the reality, where the agent may consider other products from the forward market but also the day-ahead market if its demand is not entirely covered.\\

The long-term electricity procurement problem involves the forecast of the electric energy consumption for the future period considered. In this research paper, the total quantity of electricity to be purchased over the procurement horizon is denoted $Q$. For the CAL product, this procurement horizon corresponds to a period of three years and the quantity $Q$ represents the consumption for one future year. It should be mentioned that this problem statement could be easily adapted to the case of a major electricity consumer signing a flexible bilateral contract with its retailer. In such a context, the energy price is generally set by another signal, defined in this contract, which is generally indexed on the forward prices.\\

In this research paper, the continuous trading timeline is discretised into a number of discrete time steps $t$ of constant duration $\Delta t$. In this case, the agent is assumed to be able to make only one decision per trading day, meaning that $\Delta t$ is equal to one day. In the context of the long-term electricity procurement task, a trading or procurement strategy represents the set of rules considered to make a decision. Mathematically, a procurement strategy is defined as a programmed policy $\pi: \mathcal{X} \rightarrow \mathcal{Y}$ which, based on some input information $x_t \in \mathcal{X}$ at time step $t$, outputs a trading decision $y_t \in \mathcal{Y}$ so as to maximise an objective criterion. The input, output and objective criterion considered in this research paper for the electricity procurement problem are presented in the next subsections.

\subsubsection{Procurement strategy input}
\label{SectionAlgorithmInput}

Ideally, the procurement strategy input $x_t$ at time step $t$ should encompass every single piece of information capable of affecting future electricity prices. Nevertheless, a major difficulty of the electricity procurement problem is the unavailability of such information, which can be both quantitative and qualitative, and can take various forms. This situation leads to significant uncertainty, with changes in price being impossible to accurately explain and/or predict. In this research paper, the input $x_t$ at time step $t$ is modelled as follows:

\begin{equation}
    \label{EquationInput}
    x_t = \{P_t,\ S_t\}
\end{equation}

\noindent where:

\begin{itemize}
    \item [$\bullet$] $P_t= \{p_{t-\tau}|\tau=1, ..., K\}$ is the series of $K$ previous CAL prices, $K$ being a parameter.
    
    \item [$\bullet$] $S_t$ is the trading agent state information, which is mathematically expressed as follows:
    
    \begin{equation}
        \label{EquationState}
        S_t = \{t,\ T,\ q_t,\ Q\}
    \end{equation}
    
    \noindent with:
    
    \begin{itemize}
        \item [$\bullet$] $t$ being the current trading time step.
         \item [$\bullet$] $T$ being the total number of trading time steps over the procurement horizon.
        \item [$\bullet$] $q_t$ being the quantity of electricity already purchased by the agent at time step $t$.
        \item [$\bullet$] $Q$ being the total quantity of electricity to be purchased over the procurement horizon.
    \end{itemize}
\end{itemize}

\subsubsection{Procurement strategy output}
\label{SectionAlgorithmOutput}

At each trading time step, the agent has to decide whether to purchase electricity right now or to wait for a future opportunity. Consequently, the procurement strategy output $y_t$ at time step $t$ is binary and can be mathematically expressed as the following:

\begin{equation}
    \label{EquationOutput}
    y_t \in \{0,\ 1 \}
\end{equation}

\noindent with $y_t = 0$ corresponding to the advice of waiting, and $y_t = 1$ to the advice of buying electricity.\\

Whenever purchasing electricity, the agent is required to specify the quantity traded. In this research paper, the volume contracted is assumed to be fixed. The total quantity of electricity $Q$ is simply split into $N \in \mathbb{N}\backslash\{0\}$ purchase operations of a fixed amount of electricity $A = Q/N$. Consequently, the quantity of energy purchased at each trading time step $t$ would either be equal to $0$ or $A$ depending on the algorithm output $y_t$. However, this approach does not take into account the resolution of the market $dQ$, corresponding to the smallest block of electricity tradable. To address this issue, the quantity of energy $Q$ is constrained to be a multiple of this market resolution $dQ$. Moreover, the procurement strategy parameter $N$ is constrained to be such that the amount of electricity $A = Q/N$ is a multiple of the market resolution $dQ$.\\

An important constraint is assumed regarding the procurement strategy output $y_t$. The agent is required to have purchased the exact quantity of electricity $Q$ by the end of the trading activity. Because no selling operations are permitted, the agent is not allowed to buy electricity in excess of its consumption. Moreover, anticipation is necessary as the agent is only able to buy the amount of electricity $A$ at a time. Let $n_t = (Q - q_t)/A$ be the number of remaining purchase operations to be performed by the agent at time step $t$, this quantity should never exceed the number of remaining time steps $T - t$ in practice. Eventually, this constraint is mathematically expressed as follows:

\begin{equation}
    \label{EquationConstraint}
    \sum_{t = 0}^{T} y_t \ A = Q
\end{equation}

In order to realistically simulate the trading activity associated with the electricity procurement task, the trading costs have to be considered. This research paper assumes that the only trading costs incurred by the agent are the transaction costs. As their name indicates, these costs occur when a transaction is performed. Therefore, they are modelled with a fixed fee $C_F$ to be paid per MWh of electricity purchased in the forward market.\\

Making the hypothesis that the electricity is always successfully purchased by the agent, the state variable $q_t$ is updated in line with the following equation:

\begin{equation}
    \label{EquationQuantityUpdate}
    q_{t+1} = q_t + y_t \ A
\end{equation}

\subsubsection{Objective}
\label{SectionObjective}

In the scope of the electricity procurement problem, the core objective is the minimisation of the costs incurred for buying energy. However, such an intuitive goal lacks the consideration of the risk associated with the trading activity, which should ideally be mitigated as well. In fact, there exists a trade-off between cost minimisation and risk mitigation, in accordance with the popular saying: with great risk comes great reward. However, this research paper only considers electricity cost minimisation as the objective. Therefore, the quantity to be minimised is the total cost incurred by the agent at the end of the procurement horizon $c_T$, which is mathematically expressed as follows:

\begin{equation}
    \label{EquationObjective}
    c_T = \sum_{t=0}^{T} y_t \ A \ (p_t + C_F)
\end{equation}

\subsection{Algorithm description}
\label{SectionAlgorithm}

This section thoroughly presents a novel algorithm, named \textit{Uniformity-based Procurement of Electricity} (UPE), to solve the long-term electricity procurement problem. The key idea behind this algorithm is the potential benefit to speed up or delay purchase operations with respect to a reference procurement strategy when the prices are expected to go up or down in the future. At its core, this algorithm is based on the coupling of both the identification of the dominant market direction and the estimation of the procurement uniformity level quantifying the deviation from a perfectly uniform procurement policy. The first important component of this procurement algorithm is the forecaster $F$ whose responsibility is to accurately predict the dominant market trend, either upward or downward. In this context, the trend can be defined as the general direction in which the electricity price is currently going. The forecaster $F$ takes as input a series of $K$ previous CAL prices $P_t$, which were formerly normalised, and outputs the predicted trend:

\begin{equation}
    \label{EquationForecaster}
    f_t = F(P_t)
\end{equation}

\noindent with $f_t = 1$ and $f_t = -1$ respectively corresponding to a forecast upward and downward trend.\\

A market trend is a subjective notion which possesses multiple definitions in the literature. For instance, some may argue that a surprising decrease in prices for a week is a new downward trend when others consider this behaviour as a temporary deviation within a more global upward trend lasting for months. In fact, the two opinions are right depending on the time horizon considered. This research paper adopts the following rigorous mathematical definition to eliminate any ambiguity. A smoothed version of the electricity price curve is generated by applying a lag-free low-pass filtering operation of large order $k$, typically several weeks. The resulting smoothed price at time step $t$ is mathematically expressed as follows:

\begin{equation}
    \label{EquationSmoothedPrice}
    \bar{p}_t = \frac{1}{2k+1}\sum_{\tau = t-k}^{t+k} p_{\tau}
\end{equation}

\begin{figure}
    \centering
    \includegraphics[scale=0.45, trim={1.1cm 1.5cm 2cm 1.5cm}, clip]{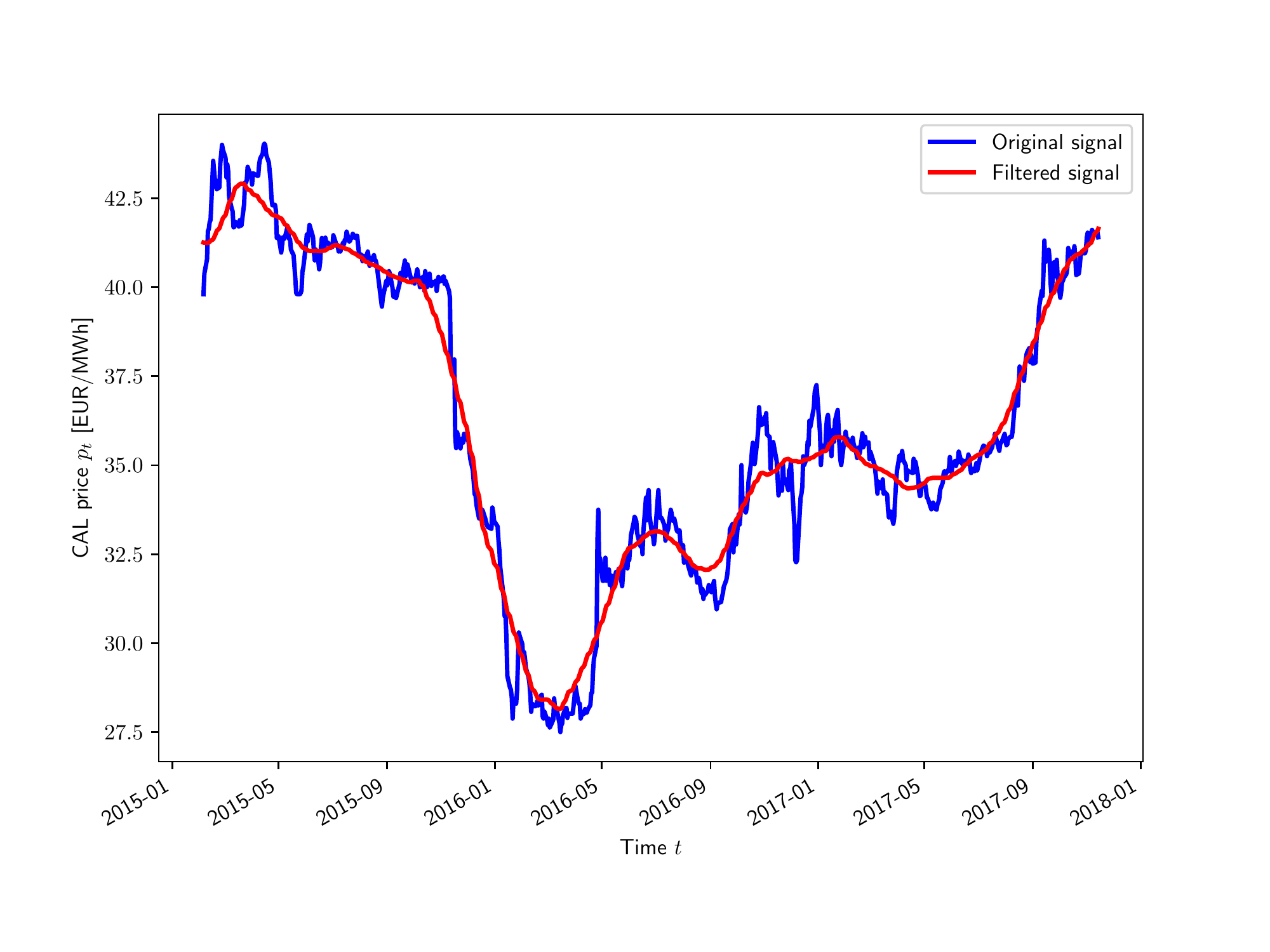}
    \caption{Low-pass filtering operation on the CAL 2018 product}
    \label{CALFiltering}
\end{figure}

As an illustration, the result of this low-pass filtering operation with $k=25$ is depicted in Figure \ref{CALFiltering} for the CAL 2018 product. The market trend at time step $t$ is defined as the difference between two consecutive smoothed prices $\bar{p}_t$ and $\bar{p}_{t-1}$. More specifically, an upward trend $\hat{f_t} = 1$ is designated when $\bar{p}_t \ge \bar{p}_{t-1}$ and a downward trend $\hat{f_t} = -1$ is specified when $\bar{p}_t < \bar{p}_{t-1}$, with $\hat{f_t}$ representing the market trend labels. This rigorous mathematical definition of a market trend is intuitive and convenient, but not perfect. As future work, more complex definitions of the market trend could be investigated. For instance, the market trend could be defined as the slope of the straight line produced by a linear regression operation on price data over a certain time period. Despite being subjective, human annotations could alternatively be considered as well.\\

The second important component of the UPE algorithm is the concept of procurement uniformity, which is based on the comparison of the current situation with a reference policy: the perfectly uniform procurement strategy. This reference policy implies buying the same amount of electricity $A_u= Q/T$ at each trading time step over the entire procurement horizon. Despite being generally not feasible in practice due to the market resolution $dQ$, this strategy is an interesting candidate for comparison purposes as the average electricity price is achieved with a risk spread over the entire procurement horizon. In doing so, this research paper introduces the procurement uniformity level $u_t \in [-1,\ 1]$ which quantifies the deviation from such a perfectly uniform strategy:

\begin{equation}
    \label{EquationTradingUniformity}
    u_t = \frac{T-t}{T} - \frac{Q-q_t}{Q}
\end{equation}

Three cases arise depending on the value of $u_t$:

\begin{itemize}
    \item [$\bullet$] \underline{$u_t = 0$:} The agent has purchased a quantity of electricity equal to the amount of energy that a perfectly uniform procurement strategy would have already bought at time step $t$.
    \item [$\bullet$] \underline{$u_t \in\ ]0,\ 1]$:} The agent is currently leading compared to a perfectly uniform procurement strategy.
    \item [$\bullet$] \underline{$u_t \in\ [-1,\ 0[$:} The agent is currently lagging compared to a perfectly uniform procurement strategy.
\end{itemize}

The decision-making process of the UPE algorithm is based on the comparison of the current procurement uniformity level $u_t$ with two trigger values $u^-$ and $u^+$. When the agent waits with purchase operations still to be performed, the procurement uniformity $u_t$ decreases over time. The idea of the proposed algorithm is to issue a new purchase operation when this indicator hits the trigger value associated with the predicted trend, $u^+$ for upward and $u^-$ for downward. Consequently, the triggers values represent how long the agent is willing to wait when a certain market trend is detected. These are parameters of the algorithm to be set by the agent according to its expectations regarding the market dynamics and its sensitivity to the trading risk.\\

\begin{figure*}
    \centering
    \includegraphics[scale=0.62, trim={0cm 0cm 0cm 0cm}, clip]{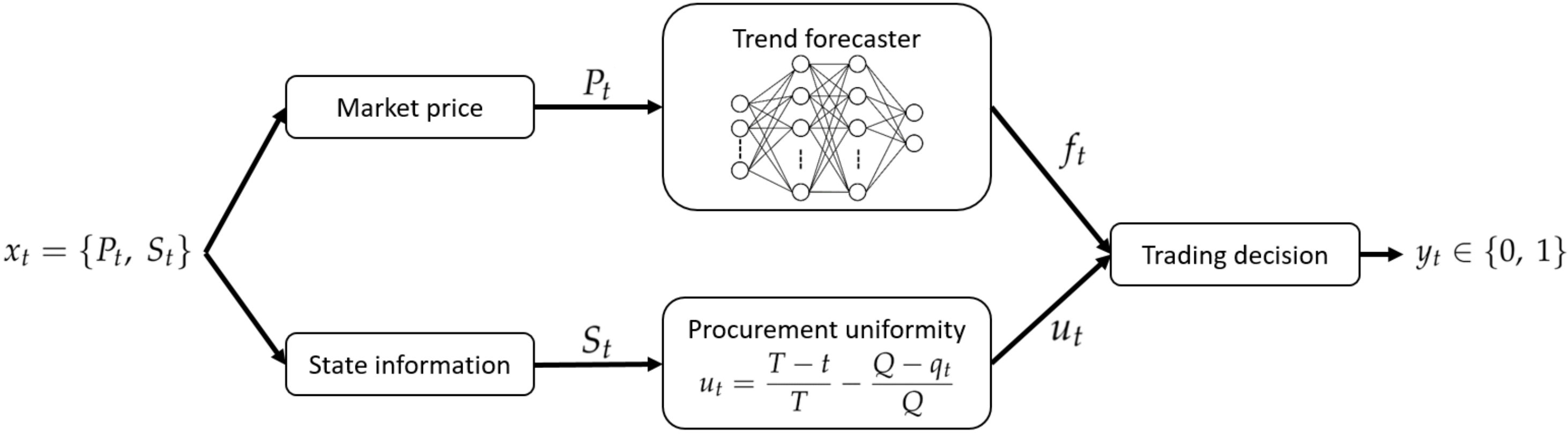}
    \caption{Flowchart of the UPE algorithm}
    \label{IllustrationUPE}
\end{figure*}

Algorithm \ref{UPE} details the decision-making process of the UPE algorithm for one time step $t$. Additionally, for the sake of clarity, a graphical illustration of the UPE algorithm is presented in Figure \ref{IllustrationUPE}. If a stable increase in the electricity prices is predicted by the forecaster $F$, happening when $f_t = 1$, the agent is instructed to wait for as long as the procurement uniformity $u_t$ remains above the trigger value $u^+$. Similarly, if a downward trend is likely to happen according to the forecaster $F$, with $f_t = -1$, the agent is advised to wait for as long as the procurement uniformity $u_t$ exceeds the trigger value $u^-$. With such a decision-making policy, the two trigger values quantify how long the agent is willing to wait when a certain trend is forecast. Consequently, $u^-$ should normally be inferior to $u^+$ as it is natural to wait longer when the prices are expected to decrease in the future.\\

\begin{algorithm*}
\caption{UPE algorithm decision-making policy for one time step $t$}
\begin{algorithmic}[1]
    \STATE \textbf{Inputs:} Procurement strategy input $x_t$, forecaster $F$ (formerly trained if necessary), trigger values $u^-$ and $u^+$.
    \STATE Execute the forecaster $f_t = F(P_t)$.
    \STATE Compute the procurement uniformity $u_t = \frac{T-t}{T} - \frac{Q-q_t}{Q}$.
    \IF{$f_t = 1$ \AND $u_t < u^+$}
        \STATE Make the trading decision to buy electricity: $y_t = 1$.
    \ELSIF{$f_t = -1$ \AND $u_t < u^-$}
        \STATE Make the trading decision to buy electricity: $y_t = 1$.
    \ELSE
        \STATE Make the trading decision to wait: $y_t = 0$.
    \ENDIF
    \RETURN $y_t$
\end{algorithmic} 
\label{UPE}
\end{algorithm*}

In this research paper, two forecasters are considered to approximate the true values of the market trend $f_t$ previously defined based on past price data only. They are respectively called \textit{Basic forecaster} and \textit{DL forecaster}. In both cases, the forecasting model is trained once on a series of past CAL prices and then remains fixed until the end of the trading activity. There is no retraining or update of the model parameters as new data progressively become available. However, this retraining/updating operation may be particularly useful in this context where the price dynamics is constantly changing. This potential improvement will be further addressed in future work.

\subsubsection{Basic forecaster}
\label{SectionBasicForecaster}

In finance, a popular approach to acquire insights about the market trend from past data consists in comparing two moving averages of different window lengths, as explained in the book \citep{Chan2009}. The idea is to assess how the more recent prices represented by the shorter moving average evolved with respect to the older prices described by the longer moving average. Both window lengths $L^{short}$ and $L^{long}$ are parameters to be tuned, with typical values being several weeks or even months. The moving average of window length $L$ for time step $t$ is mathematically expressed as follows:

\begin{equation}
    \label{EquationMovingAverage}
    M_t(L) = \frac{1}{L} \sum_{\tau = t-L}^{t-1} p_{\tau}
\end{equation}

With such a definition, an upward trend $f_t = 1$ is naturally expected when the shorter moving average is larger than the longer one, i.e. if $M_t(L^{short}) \ge M_t(L^{long})$. On the contrary, a downward trend $f_t = -1$ is awaited when the shorter moving average is smaller than the longer one, i.e. if $M_t(L^{short}) < M_t(L^{long})$. This relatively basic approach is considered for the first forecasting model $F^{MA}$ of this research paper. The UPE algorithm employing this basic forecaster is named \textit{Uniformity-based Procurement of Electricity with Moving Averages} (UPE-MA).

\subsubsection{DL forecaster}
\label{SectionAIForecaster}

A more advanced approach based on recent DL techniques is considered for the second forecasting model. This forecaster $F^{DL}$ consists of a feedforward DNN composed of $N_L$ hidden layers with $N_N$ neurons each. Leaky rectified linear unit activation functions, introduced in the article \citep{Maas2013}, are chosen for the hidden layers. Generally referred to as \textit{Leaky ReLU}, this activation function is mathematically expressed as follows:

\begin{equation}
    \label{EquationLeakyReLu}
    f(x) =  \begin{cases}
                x & \text{if } x > 0, \\
                0.01x & \text{otherwise.}
            \end{cases}
\end{equation}

\begin{figure*}
    \centering
    \includegraphics[scale=0.45, trim={0cm 0cm 0cm 0cm}, clip]{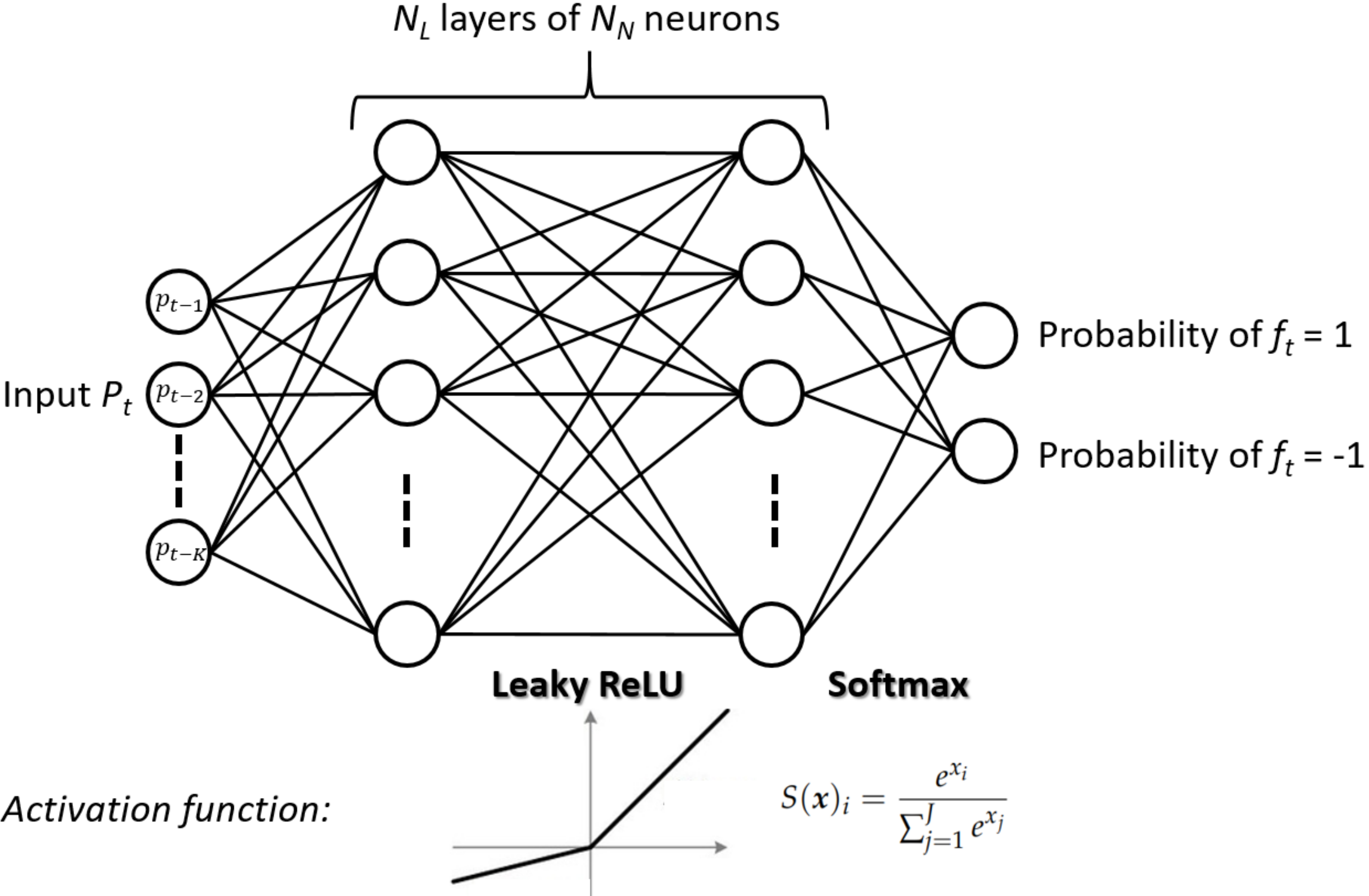}
    \caption{Illustration of the forecasting deep learning model}
    \label{IllustrationDNN}
\end{figure*}

Because the trend forecast is a classification problem, a \textit{softmax} activation function is selected for the output layer to return the probabilities associated with each trend, as explained in the book \citep{Goodfellow2016}. The forecaster $F^{DL}$ naturally outputs the market trend associated with the greatest probability. The softmax activation function takes as input a vector of real numbers $\bm{x}=(x_1, ..., x_J) \in \mathbb{R}^J$ and outputs a vector of $J$ real numbers bounded between 0 and 1 representing probabilities:

\begin{equation}
    \label{EquationSoftmax}
    S(\bm{x})_i = \frac{e^{x_i}}{\sum_{j=1}^{J} e^{x_j}}\ \ \ \ \ \forall i \in \{1, ..., J\}
\end{equation}

The training of this DNN is performed with the \textit{ADAM} optimiser introduced by the article \citep{Kingma2015} and a \textit{cross-entropy} loss to be minimised, inspired from the article \citep{Kullback1951}. Widely used for classification tasks and also referred to as logarithmic loss, the cross-entropy loss is computed as follows:

\begin{equation}
    \label{EquationCrossEntropy}
    \mathcal{L}(\theta) = \frac{1}{B}\sum_{b=1}^{B} -\log(p(y_b = \hat{y}_b|x_b, \theta))
\end{equation}

where:

\begin{itemize}
    \item [$\bullet$] $B$ is the batch size.
    \item [$\bullet$] $x$ is the DNN input.
    \item [$\bullet$] $y$ is the DNN output.
    \item [$\bullet$] $\hat{y}$ is the classification label.
    \item [$\bullet$] $\theta$ represents the parameters of the DNN.
    \item [$\bullet$] $p(\star)$ represents the probability of the event $\star$.
\end{itemize}

Additionally, both dropout and L2 regularisation techniques are adopted for generalisation purposes (for handling overfitting). All the DL techniques mentioned are covered in more details in the article \citep{LeCun2015} and the book \citep{Goodfellow2016}. Eventually, the deep learning model considered for forecasting is illustrated in Figure \ref{IllustrationDNN}. As previously suggested, the dataset used to train this DL forecasting model includes a series of previous CAL price histories $P_t$ for the inputs and a series of associated market trends $f_t$ previously defined for the outputs. In this research paper, different training, validation and test sets are considered for training the deep learning model, tuning its hyperparameters and evaluating its accuracy. The UPE algorithm operating the forecaster $F^{DL}$ is named \textit{Uniformity-based Procurement of Electricity with Deep Learning} (UPE-DL).

\section{Results}
\label{SectionResults}

This section evaluates the performance realised by the proposed UPE algorithm for the two forecasters considered. Section \ref{SectionPerformanceAssessmentMethodology} presents the performance assessment methodology. The results achieved by the UPE-MA and UPE-DL algorithms are discussed in Section \ref{SectionResultsDiscussion}.

\subsection{Performance assessment methodology}
\label{SectionPerformanceAssessmentMethodology}

\begin{figure*}
    \centering
    \includegraphics[scale=0.55, trim={3.8cm 3.3cm 4.0cm 2.7cm}, clip]{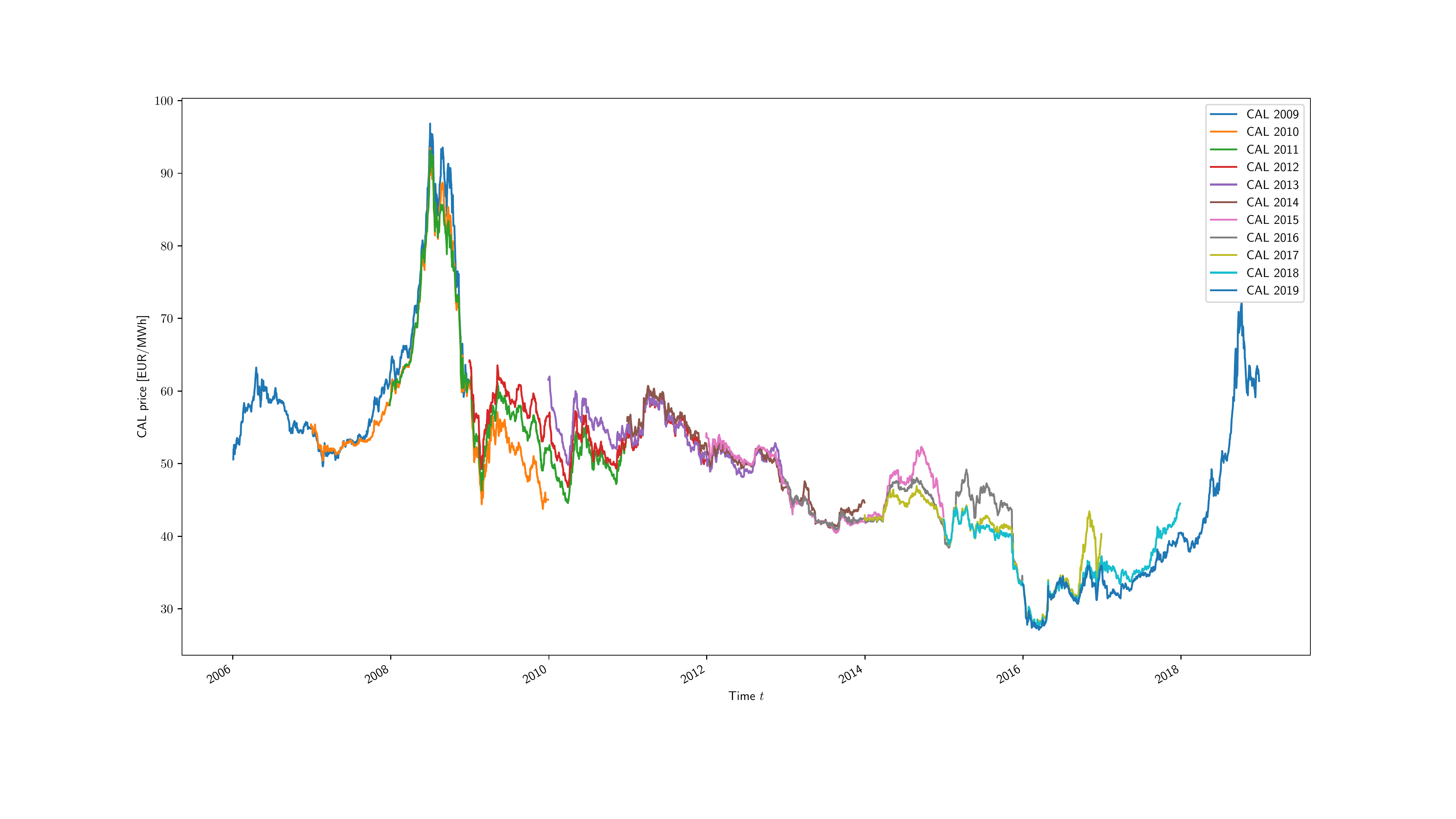}
    \caption{Illustration of the entire set of CAL data considered in the case study.}
    \label{CaseStudy}
\end{figure*}

In this research paper, the performance of a procurement strategy is evaluated on a case study consisting of CAL products over a period of eight years, ranging from CAL 2012 to CAL 2019. This enables to confront the strategy with diverse market behaviours: dominant upward and downward trends, various levels of volatility. Moreover, a clear separation between the training and test sets is imposed in order to avoid any false results due to the overfitting phenomenon. Both the tuning of the strategy parameters and the training of the DL model are performed on the CAL product three years prior to the one actually tested, so that the training and test sets do not share any data. For instance, the training of a procurement strategy for the CAL 2018 product, with electricity purchased between 2015 and 2017, is performed on the CAL 2015 one, with energy bought between 2012 and 2014. For the sake of clarity, the entire set of CAL data considered in the case study is depicted graphically in Figure \ref{CaseStudy}. If missing data or abnormal values are detected, new samples are generated by linear interpolation or extrapolation as replacements for the problematic samples. One can also point out that a CAL product consists of approximately 750 daily prices.\\

For comparison purposes, two basic benchmark procurement strategies are considered in this research paper. The first one is named \textit{Naive Balanced Electricity Procurement} (NBEP). This strategy simply consists in dividing the procurement horizon into $N$ intervals of identical durations, and executing a purchase operation in the middle of each interval. The second benchmark procurement strategy is named \textit{Electricity Procurement with Moving Averages} (EPMA), and is an adaptation of the popular moving averages trend following strategy to the electricity procurement task. More details about this specific trading strategy widely used in the stock markets can be found in the book \citep{Chan2009}. The resulting algorithm is based on the same principle as the basic forecaster presented in Section \ref{SectionBasicForecaster}, with two moving averages of different durations for estimating the market trend. A purchase operation is triggered each time a new upward trend is predicted, occurring when the shorter moving average $M_t(L^{short})$ crosses and becomes higher than the longer moving average $M_t(L^{long})$. If the number of purchase operations performed with this policy is smaller than $N$ by the end of the procurement horizon, the remaining ones are executed at the last trading time steps.\\

As previously explained in Section \ref{SectionObjective}, the procurement strategy objective is the minimisation of the total cost $c_T$. To improve the readability of the results, the quantitative performance indicator $C = c_T / Q$, representing the average price expressed in \euro{}/MWh at which the electricity is purchased, is considered instead. Moreover, several reference procurement policies achieving benchmark values for this quantitative performance indicator are considered for comparison purposes. Firstly, the best and worst procurement strategies achieving respectively the minimum and maximum values for the indicator $C$ are examined. They are trivially computed once all the prices for the entire trading horizon are known. Secondly, the mean electricity price achieved by a perfectly uniform procurement strategy is computed, although this policy is generally not feasible in practice due to the market resolution $dQ$. Lastly, the UPE algorithm equipped with an ideal forecaster achieving 100\% accuracy, i.e. always correctly predicting the trend labels $\hat{f_t}$ defined in Section \ref{SectionAlgorithm}, is considered under the name UPE-F.

\subsection{Results analysis}
\label{SectionResultsDiscussion}

For the reproducibility of the results presented in this section, Table \ref{Parameters} reveals the hyperparameters used in the simulations. Additionally, the case study data depicted in Figure \ref{CaseStudy} are provided by Ice Endex \citep{DataCAL}. This set of data is not freely and publicly available, but can be purchased or potentially requested for research purposes. In accordance with the performance assessment methodology, Table \ref{TablePerformance} presents the results achieved by both the benchmark (NBEP, EPMA) and proposed (UPE-MA, UPE-DL) procurement strategies, together with the reference policies.\\

\begin{table}
\small
  \caption{Hyperparameters used in the simulations.}
  \label{Parameters}
  \centering
  \begin{tabular}{lcc}
    \toprule
    \textbf{Name} & \textbf{Symbol} & \textbf{Value}\\
    \midrule
    Number of days in input variable $P_t$ & $K$ & 50 \\
    Low-pass filtering operation order & $k$ & 25 \\
    Quantity of electricity to buy [MWh] & $Q$ & 100000 \\
    Number of purchase operations & $N$ & 10 \\
    Procurement uniformity trigger $-$ & $u^-$ & $-0.3$ \\
    Procurement uniformity trigger $+$ & $u^+$ & 0 \\
    Number of layers in the DNN & $N_L$ & 5 \\
    Number of neurons per layer & $N_N$ & 1024 \\
    Dropout probability & $D_p$ & 0.2 \\
    L2 factor & $L_2$ & $10^{-4}$ \\
    ADAM learning rate & $l_r$ & $10^{-6}$ \\
    Number of epochs & $n$ & $30000$ \\
    \bottomrule
  \end{tabular}
\end{table}

\begin{table*}
  \caption{Comparison of the electricity cost $C$ achieved by the procurement strategies.}
  \label{TablePerformance}
  \centering
  \begin{tabular}{ccccccccc}
    \toprule
    \multirow{2}{*}{\textbf{CAL product}} & \multicolumn{4}{c}{\textbf{Procurement strategies}} & \multicolumn{4}{c}{\textbf{References}}\\
    \cmidrule(r){2-5}
    \cmidrule(r){6-9}
    & \textbf{NBEP} & \textbf{EPMA} & \textbf{UPE-MA} & \textbf{UPE-DL} & \textbf{Min} & \textbf{Mean} & \textbf{Max} & \textbf{UPE-F} \\
    \midrule
    2012 & 54.903 & 52.854 & 56.076 & \textbf{52.032} & 47.289 & 55.005 & 63.319 & 52.879 \\
    2013 & 53.564 & \textbf{52.566} & 52.826 & 53.653 & 48.239 & 53.614 & 60.638 & 52.400 \\
    2014 & 49.387 & 51.346 & 50.063 & \textbf{48.762} & 41.233 & 50.234 & 60.279 & 49.036 \\
    2015 & 45.834 & \textbf{44.402} & 47.400 & 47.656 & 40.580 & 47.043 & 53.713 & 46.230 \\
    2016 & 43.613 & \textbf{43.038} & 43.927 & 43.374 & 34.077 & 43.912 & 48.631 & 41.632 \\
    2017 & 38.887 & \textbf{38.707} & 39.940 & 39.477 & 27.814 & 39.449 & 46.435 & 37.756 \\
    2018 & 36.357 & 38.666 & \textbf{34.537} & 36.251 & 27.727 & 36.440 & 43.852 & 35.443 \\
    2019 & 42.532 & 48.526 & 37.749 & \textbf{37.501} & 27.310 & 39.017 & 70.693 & 37.335 \\
    \midrule
    \midrule
    \textbf{Average} & 45.635 & 46.263 & 45.315 & \textbf{44.838} & 36.784 & 45.589 & 55.945 & 44.089 \\
    \bottomrule
  \end{tabular}
\end{table*}

\noindent \textbf{Average performance:} Considering only the last line of Table \ref{TablePerformance}, the two variants of the UPE algorithm outperform both benchmark procurement strategies on average. Moreover, the UPE-MA and UPE-DL algorithms respectively perform 0.6\% and 1.65\% better than a perfectly uniform procurement strategy (reference policy \textit{Mean}). This figure corresponds to the average relative reduction in electricity cost $C$ between two procurement strategies for the case study considered. It indicates that the UPE algorithm is able to correctly identify and exploit certain market phenomena. This also suggests that the forecaster $F^{DL}$ outputs more accurate market trend predictions compared to the more basic forecaster $F^{MA}$, the accuracy of the forecaster $F$ being defined as the number of correct predictions $f_t = \hat{f_t}$ over the total number of predictions. This interpretation is backed up by the UPE-F policy which achieves a 100\% accuracy and realises an even better average performance. Although the improvement in performance achieved by the proposed algorithm may appear to be quite limited at first glance, it corresponds to a comfortable annual saving of tens or even hundreds of thousands of euros for major electricity consumers/retailers. For instance, the UPE-DL strategy achieves on average a yearly saving of \euro{}75,100 with respect to the perfectly uniform procurement strategy for an annual consumption of 100 GWh of electricity.\\

\noindent \textbf{Results variance:} As indicated in Table \ref{TablePerformance}, the mean electricity price significantly varies over the years. Therefore, the variance of the procurement strategy performance should be assessed after subtracting this mean electricity price from the achieved electricity cost $C$. The results variance substantially differs depending on the procurement policy considered. On the one hand, the EPMA strategy achieves the best results for half of the years considered within the case study, but totally fails the CAL 2019 product due to a flaw in its design (explained at the end of this section). On the other hand, the NBEP strategy is never the best procurement policy but achieves a lower variance without any significant failure. Concerning the UPE algorithm, both variants deliver consistent results which are at least comparable and generally better than the reference mean electricity price. This consistency throughout the years demonstrates the stability of the UPE algorithm, this property being defined as the ability to generate positive results whatever the price dynamics. The stability of a procurement strategy is particularly important for the electricity procurement problem owing to considerable uncertainty. The non-negligible variance observed in Table \ref{TablePerformance} also highlights the intended diversity of the case study, with multiple market phenomena handled better or worse by each procurement policy.\\

\begin{figure*}
    \centering
    \includegraphics[scale=0.55, trim={2.7cm 2.1cm 3.2cm 2.1cm}, clip]{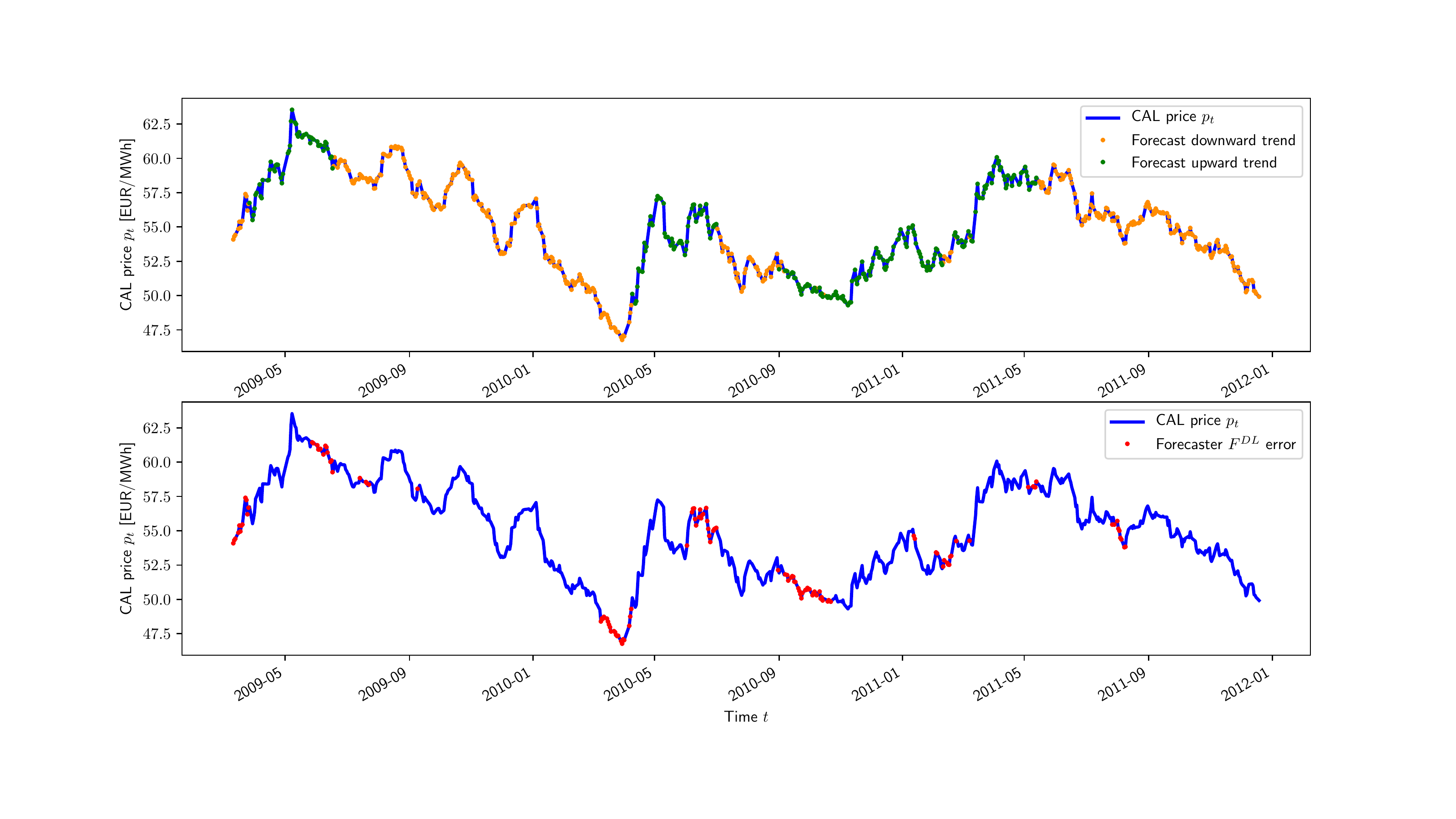}
    \caption{Forecasting model $F^{DL}$ output (top) and forecasting errors (bottom) for the CAL 2012 product.}
    \label{2012Forecasts}
\end{figure*}

\begin{figure*}
    \centering
    \includegraphics[scale=0.55, trim={2.6cm 2.1cm 3.2cm 2.1cm}, clip]{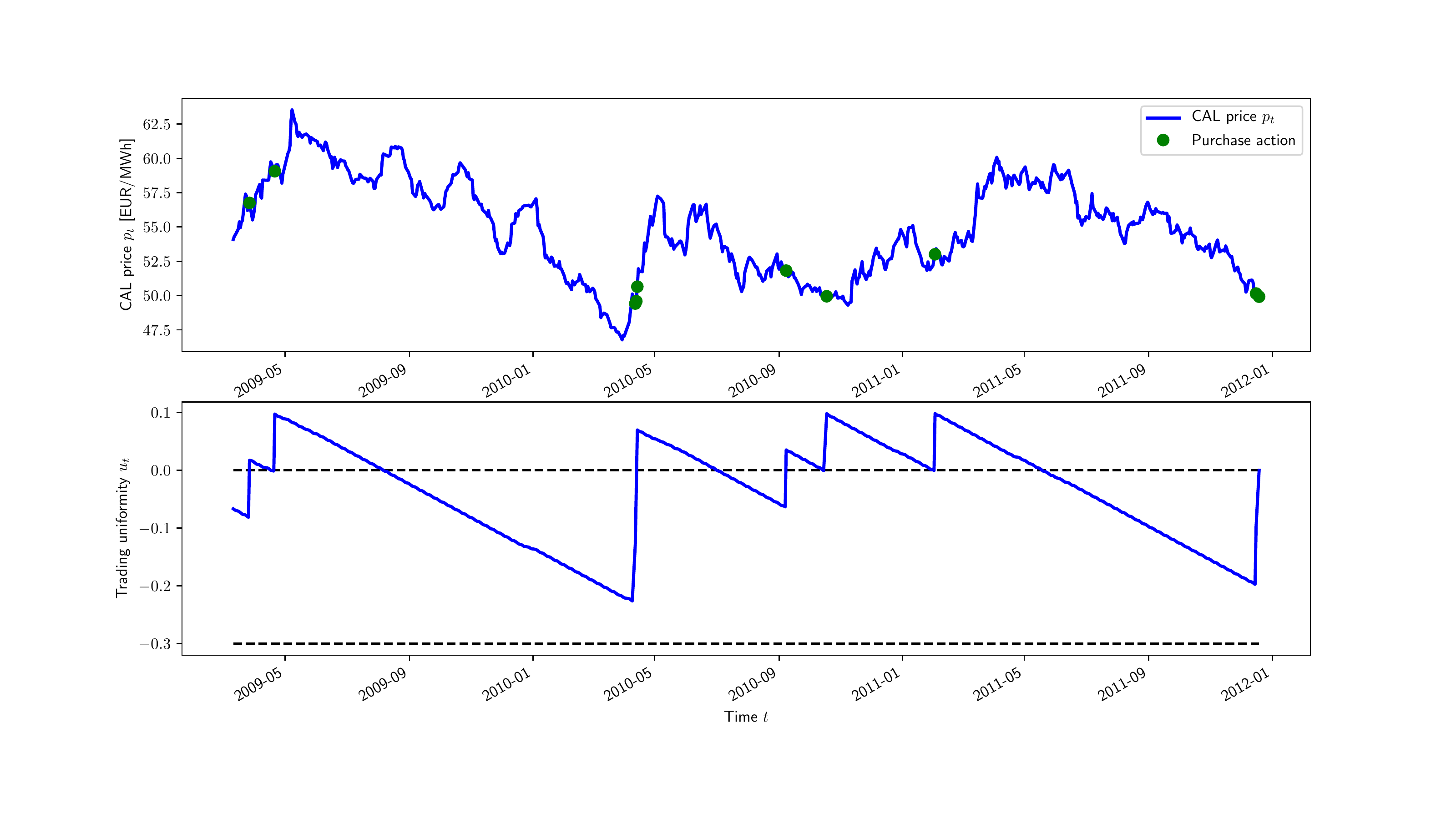}
    \caption{Purchase operations executed by the UPE-DL algorithm (top) and procurement uniformity $u_t$ (bottom) for the CAL 2012 product.}
    \label{2012Rendering}
\end{figure*}

\noindent \textbf{Typical execution of the UPE-DL algorithm:} Figures \ref{2012Forecasts} and \ref{2012Rendering} illustrate the execution of the top-performing UPE-DL procurement strategy for the CAL 2012 product. Firstly, Figure \ref{2012Forecasts} presents the predictions $f_t$ outputted by the forecaster $F^{DL}$ together with the electricity price $p_t$ in the upper plot, and the forecasting errors $f_t \neq \hat{f_t}$ in the bottom plot. For this particular year, the DL forecasting model achieves an encouraging accuracy of approximately 80\%. Moreover, the predictions do not incorrectly oscillate between the two market trends during periods of pronounced volatility, a behaviour which could significantly harm the performance of the UPE-DL algorithm. For instance, if an important downward trend occurs and if an upward trend is wrongly predicted several times during a short temporary rebound in prices, some purchase operations may be triggered too early at a higher price. Secondly, Figure \ref{2012Rendering} depicts the electricity price $p_t$ evolution together with the purchase decisions $y_t = 1$ in the upper plot, and the associated procurement uniformity level $u_t$ in the bottom plot. This figure illustrates the ability of the UPE-DL procurement strategy to delay purchase operations when the prices are expected to decrease in the future, so that they are executed close to local minima. Figures \ref{2012Forecasts} and \ref{2012Rendering} also demonstrate the interpretability of the trading decisions outputted by the UPE-DL algorithm. The fact that these decisions are easily explainable and completely transparent to the human supervisor (as opposed to a black box model) really eases the monitoring of the procurement strategy and improves its reliability, a feature which is very important in practice for the industry.\\

\begin{figure}
    \centering
    \includegraphics[scale=0.63, trim={0.9cm 0.3cm 1.4cm 1.3cm}, clip]{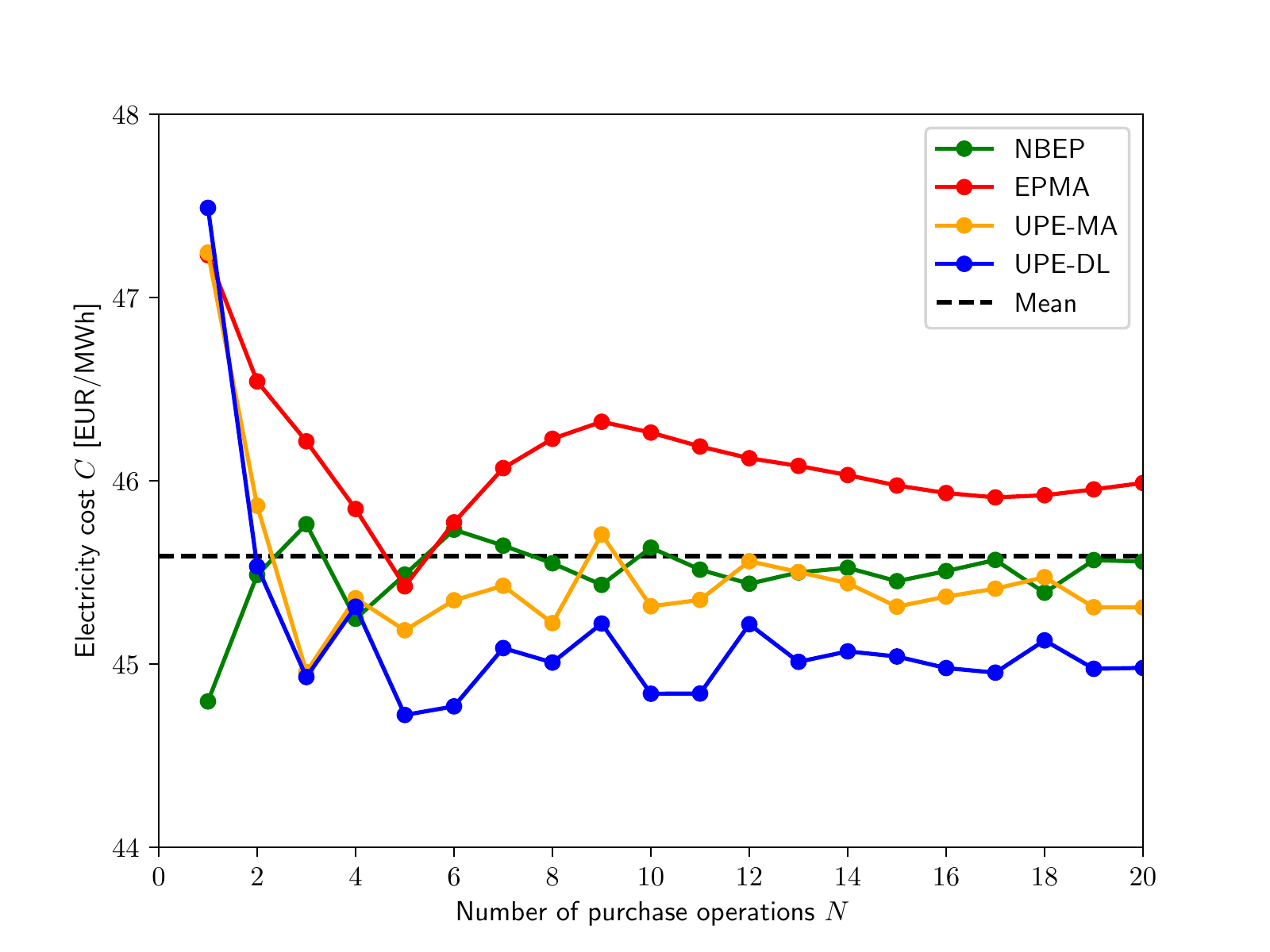}
    \caption{Effect of the number of purchase operations $N$ on the electricity cost $C$ achieved by the procurement strategies on average.}
    \label{sensitivityAnalysisN}
\end{figure}

\noindent \textbf{Sensitivity analysis:} The long-term electricity procurement problem depends on the number of purchase operations $N$ to be performed over the procurement horizon. Contrarily to the other parameters which can be freely tuned, this quantity is constrained due to the market resolution $dQ$. This is why it is important to assess the sensitivity of the number of purchase operations on the performance of the procurement strategies. Figure \ref{sensitivityAnalysisN} depicts the sensitivity of the electricity cost $C$ achieved by each procurement strategy on average with respect to this parameter $N$. Firstly, it can be observed that the UPE-DL algorithm is the leading strategy by a reasonable margin when $N > 4$. When the number of purchase operations is too limited, this is a completely different situation which should be avoided because the performance of each procurement strategy is generally the result of luck. Secondly, the UPE-DL curve is roughly shifted down compared to the UPE-MA one, which is once more an indication that the DL forecasting model improves the market trend predictions. Thirdly, larger values for the parameter $N$ may be favoured as both the UPE-MA and UPE-DL algorithms performances stabilise when the number of purchase operations increases. Regarding the benchmark procurement strategies, the NBEP one is resilient to a change in the parameter $N$ by design, and its performance tends toward the mean electricity price when this parameter increases. On the contrary, the EPMA strategy does not monitor the number of remaining purchase operations (flaw in its design). The policy executes a fixed number of purchase operations which is dependent on the number of market trend inversions, and it executes the remaining purchase operations at the end of the procurement horizon. This may lead to an unacceptable behaviour for large values of the parameter $N$, especially when the last prices are high compared to the average electricity price.

\section{Discussion}
\label{SectionDiscussion}

As previously explained in Section \ref{SectionLiteratureReview}, the sequential decision-making problem behind the electricity procurement task is formalised in many different ways in existing scientific literature (time horizons, electricity power sources and markets, electricity consumption). This means that making a fair comparison between these solutions is not really feasible. However, it is still possible to highlight the novel aspects of the proposed approach because it differs considerably from the solutions to be found in current literature. Firstly, this new method manages to successfully exploit the forecasting capabilities of recent deep learning techniques to make sound trading decisions for the long-term procurement task. To the authors' knowledge, this is the first work which considers advanced artificial intelligence techniques to solve the sequential decision-making problem behind electricity procurement, as previous solutions were mainly based on stochastic programming and optimisation techniques. Secondly, the proposed method presents the key advantage of outputting trading recommendations which are easier to understand and explain, this feature being particularly important for the industry. The reason for this is the understandable intuition from a the human perspective regarding the decision-making process, i.e. coupling trend forecasting with procurement uniformity. Thirdly, the new method developed is characterised by an automatic risk mitigation mechanism thanks to the limited deviation with respect to a perfectly uniform procurement policy (according to the procurement uniformity triggers $u^-$ and $u^+$). This makes the resulting procurement policy far more robust to exceptional events such as economic crises which may lead to the forecasting model being relatively inaccurate. Indeed, a deep learning forecasting model is expected to be quite unpredictable in such situations since the data from the test set may significantly differ from the data included in the training set. This could lead to poorly performing policies if left untouched. However, this potential collapse in performance can be prevented, or at least mitigated, thanks to the constraint on the deviation with respect to a perfectly uniform procurement policy, if the tuning of the two parameters $u^-$ and $u^+$ is appropriate.\\

Regarding the weaknesses, the proposed method suffers from two main limitations. Firstly, the quality and relevance of the trading recommendations outputted by the UPE algorithm are strongly dependent on the performance (accuracy) of the trend forecasting module. However, if a complex deep learning model is considered for that purpose, the forecaster $F$ becomes a black box model which may be quite difficult to properly analyse. This may have a negative impact on the ability to explain these trading recommendations. Secondly, as previously explained, the deviation from a perfectly uniform procurement policy is restricted according to the two procurement uniformity triggers $u^-$ and $u^+$ (tunable parameters). Although this mechanism may be particularly useful to effectively mitigate the trading risk, it may also limit the profit (reduction in costs) achievable when the forecasting module is very accurate. As a consequence, this particular mechanism can be considered both as a strength and as a weakness, depending on the market dynamics and on the trend forecaster performance. In addition, the tuning of the two parameters $u^-$ and $u^+$ is an appreciable degree of freedom, but which may also be quite tricky to perform.\\

Even though the UPE-DL algorithm achieves promising results, several avenues to further improve the solution exist. Firstly, the deep learning forecasting model could be significantly improved by following the recommendations of the scientific literature. For instance, long short-term memory layers, which are introduced in the article \citep{Hochreiter1997}, could be considered as they have already proven to better process time-series data at the cost of increased complexity. One could also consider retraining/updating the forecasting model at a certain frequency as new data progressively become available, in order to constantly adapt to current market dynamics. Other state of the art forecasting techniques may also be considered, please refer to articles \citep{Sezer2020} and \citep{Fawaz2019} for more information. Secondly, the procurement strategy input $x_t$ is not sufficient to accurately explain all market phenomena observed in the case study. Other information such as macroeconomic data, correlated commodities prices, or news should be factored into the input $x_t$ to improve the accuracy of the forecaster. Thirdly, the trading risk associated with the long-term electricity procurement problem should be mathematically defined. Once properly quantified, this risk should be considered in the objective of the procurement strategies together with cost minimisation. Lastly, novel deep reinforcement learning techniques could be well-suited to solve the complex decision-making problem behind the long-term electricity procurement task. This approach should be considered in the future, drawing on what the article \citep{Theate2020} realised for another algorithmic trading problem.\\

To end this section, the feasibility of deploying the proposed solution in a real-life production environment is discussed. In fact, the present scientific article is the result of research conducted to solve the real long-term electricity procurement task of a major energy consumer. This actor was willing to take a step towards automating its procurement of electricity in order to reduce its energy costs. To be acceptable, the solution would have to be relatively easy to monitor on a daily basis by inexperienced employees. Moreover, the trading recommendations outputted by the algorithm would have to be motivated, or at least easy to understand for these employees, meaning that the reasons for doing a certain trading action should be completely transparent. The solution presented in this research paper is intended to be shortly deployed in the real production environment of this major consumer of electricity. Based on this successful experience, it can be concluded that it is perfectly feasible to deploy this solution in a real-life production environment. The only missing piece is a complete software solution to collect the required data, run the decision-making algorithm and interact with the final user. Although this new algorithmic solution has a cost (hardware, software and data), it can be viewed as a long-term investment that becomes profitable after a certain number of years depending on the electricity consumption of the consumer or retailer.

\section{Conclusions}
\label{SectionConclusions}

The present scientific research paper proposes a novel algorithm, named \textit{Uniformity-based Procurement of Electricity} (UPE), advising a retailer or a major consumer of electricity for its procurement task in the forward market, especially for the CAL product (Belgian Power Base Futures). This algorithm relies on a forecasting mechanism to predict the market trend and on the concept of procurement uniformity, which quantifies the deviation from a perfectly uniform reference policy purchasing a tiny amount of energy at each time step over the entire procurement horizon. Two variants of this algorithm were developed depending on the forecasting model considered, respectively UPE-MA for moving averages and UPE-DL for deep learning. On average, both variants surpass the benchmark procurement strategies, and the top-performing UPE-DL algorithm achieves a reduction in costs of 1.65\% with respect to a perfectly uniform policy achieving the mean price. This represents an average yearly saving of \euro{}75,100 for an annual consumption of 100 GWh of electricity between 2012 and 2019. Moreover, the approach presented in this research paper exhibits key advantages. Firstly, the algorithm is relatively stable, with consistent results achieved throughout the years despite the difficulty of having to deal with various market phenomena. Moreover, the decision-making process is sufficiently robust with respect to exceptional events, such as economic crises, thanks to the limited deviation with respect to a perfectly uniform procurement policy mechanism. Secondly, the decisions advised by this solution are totally transparent and easily explainable, which is key for the industry and improves the reliability of the procurement strategy. Thirdly, the proposed methodology could be slightly adapted to address the sequential decision-making problem of selling electricity in the forward market, which could be particularly useful for energy producers. Lastly, the approach presented in this scientific article is general and may be well suited to solve other commodity procurement problems presenting similar constraints.

\section*{Acknowledgments}

Thibaut Th\'{e}ate is a Research Fellow of the F.R.S.-FNRS, of which he acknowledges its financial support.

\bibliography{Article.bib}

\end{document}